\documentclass[aps,reprint,superscriptaddress,
nofootinbib,
amsmath,amssymb,
prc
]{revtex4-2}
\usepackage{natbib}
\usepackage{graphicx,color}
\usepackage{newtxtext}
\usepackage{dcolumn}
\usepackage{bm}
\usepackage[T1]{fontenc}

\graphicspath{{./}}
\usepackage{url}
\usepackage[colorlinks=true,allcolors=blue,breaklinks]{hyperref}
\usepackage[normalem]{ulem}

\begin{document}

\preprint{...}

\title{Exploring medium mass nuclei using effective chiral nucleon-nucleon interactions}
\author{Sota Yoshida}
\email{syoshida@cc.utsunomiya-u.ac.jp}
\affiliation{Institute for Promotion of Higher Academic Education,~Utsunomiya University,~Mine,~Utsunomiya,~321-8505,~Japan}

\date{\today}

\begin{abstract}
In the past decades, it has been shown that the three-body force is necessary for a fully microscopic understanding of nuclear many-body systems, and thus efficient schemes for storing and utilizing three-body matrix elements have been developed.
However, three-body matrix elements still demand a lot of memory and/or storage and it readily exceeds the limitation of the state-of-the-art supercomputers.
We explore the possibility of mimicking the effects of three-body forces in medium-mass region, especially lower $pf$ shell, by density-dependent effective nucleon-nucleon interactions combining with a modern family of chiral nucleon-nucleon interaction and an uncertainty quantification on the coupling constants.
We found that some effective nucleon-nucleon potentials give decent description of ground state energies, but also systematic underestimation in charge radii and deficiencies relevant to the $pf$-shell region.
\end{abstract}

\maketitle

\paragraph*{Introduction.}

Chiral effective field theory (EFT)~\cite{EMrev1,EMrev2,LENPICrev1,LENPICrev2} provides a systematic and hierarchical description of nucleon-nucleon (NN) force and three-nucleon force (3NF), and has become a standard approach for microscopic nuclear structure and reaction studies.
The potentials are parametrized in terms of low-energy constants (LECs), and the LECs at NN-sector are determined to reproduce e.g. scattering data. 

While many applications of chiral potentials so far were usually based on the NN force up to next-to-next-to-next-to-leading order (N3LO) plus the 3NF up to next-to-next-to-leading order (NNLO), the NN part (and sometimes 3NF too) in structure and reaction calculations is gradually being replaced by higher order one such as next-to-next-to-next-to-next-to-leading order (N4LO)~\cite{EMN}, see e.g.~\cite{HutherN4LO,KravvarisN4LO,Kravvaris2020} for recent applications.

In this work, we explore the capability of effective NN interactions, consisting of N4LO NN force and density-dependent effective NN force from 3NF at NNLO~\cite{HoltKaiserWeisse,Kohno,*KohnoErratum} without introducing explicit 3NFs at all.
Although it is a drastic attempt, this should make it easier, in terms of computational burden, to explore the structure of various nuclei across the nuclear chart using microscopic many-body methods such as the post-Hartree-Fock methods and configuration interaction methods.
The main focus of this paper is on a specific intermediate-mass region, in particular a lower $pf$-shell region, as the testing ground. We found that our effective NN interactions are qualitatively well, and still need some improvements for more quantitative discussions.
We have also introduce a newly developed Julia package covering chiral potentials and various many-body methods for nuclear structure calculations. The package enables us more flexible studies such as uncertainty quantification in medium-mass regions.



\paragraph*{Chiral potentials.}

For the NN part, we use the so-called Entem-Machleidt-Nosyk (EMN) interaction~\cite{EMN} up to N4LO with cutoffs $\Lambda=500$ MeV and $\Lambda_\mathrm{SFR}=700$ MeV, where $\Lambda_\mathrm{SFR}$ is a cutoff for the spectral function regularization~\cite{Epelbaum2004} for two-pion exchange (TPE) contributions. 
Detailed expressions of the TPE contributions are described in Ref.~\cite{EKMN}.
Then, we soften the NN interaction with the similarity renormalization group (SRG) in the momentum space to get better convergence of binding energies with respect to the model space size, which is typically specified by $e_\mathrm{max}$ truncation for harmonic oscillator quanta $e=2n+l \leq e_\mathrm{max}$.

Besides, we attempt to approximate both induced and genuine 3NFs by a density-dependent NN force. We refer to a density-dependent NN force from three-body forces as "2n3n" to distinguish it from NN and induced/genuine 3NF.
Since the detailed derivation of the 2n3n contribution at NNLO can be found in
Refs.~\cite{HoltKaiserWeisse,Kohno,*KohnoErratum}, we do not repeat the derivation here.
Note that this approximation and the matrix elements have been partly used in our previous study~\cite{NTsunoda20} to investigate the impact of the choice of three-body forces on the neutron driplines around $Z=9$--$12$.

As discussed in Ref.~\cite{EMN} and references therein, the TPE contribution in the 3NF at N3LO and N4LO have essentially the same mathematical structure at NNLO and thereby can be approximated in terms of effective LECs at NNLO, i.e., $c_1,c_3,c_4$. Therefore, we regard the LECs for TPE contribution in the 3NF as independent parameters, and refer to them as $\tilde{c}_1, \tilde{c}_3, \tilde{c}_4$ to distinguish from the ones that appear in NN sector.
The other LECs for 3NF appear at NNLO, i.e., the one-pion exchange term $c_D$ and contact term $c_E$.
The 2n3n terms are also dependent on the Fermi momentum $k_F$, and its dependence on target nuclei is nontrivial. For this reason, we fix it as $k_F = 1.35$ $\mathrm{fm}^{-1}$ (corresponding to an empirical normal nuclear density $\rho = 0.166$ $\mathrm{fm}^{-3}$), and restrict our considerations to the nuclei having similar mass numbers and binding energy per nucleon, $^{40,48,52,54}$Ca and $^{56}$Ni, and use them to fix the LECs.

Throughout this study, we work in the two-body space only, expecting both induced 3NF via SRG and genuine 3NF can be mimicked by the 2n3n approximation. If it works quantitatively well, this can be one solution to the problem of computational costs for three-body matrix elements. If not, it indicates that the structure of 3NF is much richer than that of a density dependent NN interaction at NNLO and one needs to consider sub-leading contributions or three-body matrix elements rather explicitly. One of the aims of this work is to consider and clarify the capability of such effective NN potentials in medium mass regions using a modern chiral NN potential up to N4LO.

To summarize, we work with both NN and 2n3n contributions up to N4LO (in an effective viewpoints for the latter), and the free parameters in our potential to be calibrated below are the five LECs for the 2n3n terms, $\vec{c}=\{\tilde{c}_1, \tilde{c}_3, \tilde{c}_4, c_D, c_E\}$.

\paragraph*{Many-body methods.}
In this work, we adopted a couple of many-body methods to calculate medium-mass nuclei,
and all the methods below are originally implemented in NuclearToolkit.jl~\cite{NuclearToolkit}.
One is the so-called Hartree-Fock many-body perturbation theory (HF-MBPT).
This gives faster evaluations for g.s.~properties of medium-mass nuclei and thus will be used to calibrate LECs for the 2n3n.
With the HF-MBPT, one can evaluate perturbative corrections to the estimation of observables at the HF level, known as M{\o}ller-Plesset perturbation theory in the field of chemistry.
More details can be found in Refs.~\cite{ShavittBartlett2009,TichaiRev,Miyagi22Sn} on which our implementation is based.

Another one is the in-medium similarity renormalization group (IMSRG)~\cite{Tsukiyama2011}.
The calculation is performed with the Magnus formulation of IMSRG~\cite{Morris_Magnus} and the arctangent generator~\cite{Hergert2016Rev}.
We truncate the IMSRG flow equation up to the normal-ordered two-body level at each step, which is the so-called IMSRG(2) truncation. The comparisons with HF-MBPT and IMSRG(2) will be discussed later.

In addition to these, the valence-space IMSRG (VS-IMSRG)~\cite{StrobergRev19} is applied to derive shell-model effective interactions for a target model space. The generator to decouple between valence space and core/outer space is again the arctangent generator. More precisely, we employ a more generalized one with the denominator delta $\Delta$ introduced in Ref.~\cite{Miyagi20} to decouple a multi-shell valence space. However, $\Delta$ is now set as 0.
Valence shell-model calculations are performed using both NuclearToolkit.jl and KSHELL~\cite{KSHELL1,*KSHELL2}.
The latter is more appropriate for larger systems.

\paragraph*{NuclearToolkit.jl.}
We developed a new Julia package for nuclear structure calculations, named NuclearToolkit.jl~\cite{NuclearToolkit}, covering chiral EFT potentials, HF-MBPT, IMSRG/VS-IMSRG, and the valence shell model.
The shell-model part of this package had been developed as an independent code named ShellModel.jl, and used in the previous study~\cite{SY_EC} to demonstrate the shell-model emulator with the eigenvector continuation.
The package is neither a wrapper nor a Julia translation of existing codes such as EM potential~\cite{EMrev1}, imsrg~\cite{imsrgcode}, KSHELL~\cite{KSHELL1,*KSHELL2}, etc., while these pioneering works provide us with a certain guideline to develop NuclearToolkit.jl.
The package is designed as an open source software for reproducibility and transparency of upcoming works
and for further acceleration of international research collaborations in the community.
It is automated by GitHub Actions to generate the document from {\it docstring} on top of each function, deploy the document, and run the test code to avoid destructive changes. 

With the help of features of Julia language~\cite{Julia} (high readability, performance, portability, package manager, etc.), one can use the package on a laptop or remote environment without root permissions in the almost same way.
The author would be pleased if this package could promoted open science, and served as a basis for exciting studies by readers.

\paragraph*{Calibration of LECs for the 2n3n contributions.}
Next, we explain our procedures to calibrate the five LECs for the 2n3n terms.
The aim of this work is to explore the capability of a density-dependent NN interaction
to spare discrepancy between the NN-only results and the nature without introducing explicit 3NFs.
Parameter calibration and uncertainty quantification in the following will not be exhaustive to examine, e.g., dependences on the priors and adopted data, but enough for the purpose.
Our strategy can be divided into the following three steps and more technical details are described later:
(i).~Defining a prior distribution to encode the so-called naturalness of LECs and a likelihood function for target nuclei. (ii).~Carrying out Markov chain Monte Carlo (MCMC) samplings over the five-dimensional space with relatively smaller model spaces $e_\mathrm{max}=6,8$, $\lambda_\mathrm{SRG}=2.0$ fm$^{-1}$, and the oscillator parameter $\hbar\omega=20$ MeV.
(iii).~Evaluating an optimal value for $e_\mathrm{max}\geq 12$ based on the maximum a posteriori (MAP) value and correlation seen in the results with smaller $e_\mathrm{max}=6,8$.

For step (i), it is expected that the g.s.~energies of the five nuclei listed above are not enough to constrain all the degrees of freedom among the five LECs, while carrying out a sampling with many observables is intractable.
Therefore, we adopted independent Gaussian priors for the five LECs to softly encode the naturalness and the deviations of $\tilde{c}_1,\tilde{c}_3,\tilde{c}_4$ from $c_1,c_3,c_4$ are imposed to be smaller. The log-prior can be written as
\begin{align}
\log Pr(\vec{c}) = - \sum^5_{i=1} \frac{(c_i -\mu_i)^2}{2\sigma^2_i}.
\label{eq:prior}
\end{align}
The mean values $\{ \mu_i \}$ are $c_1,c_3,c_4$ of EMN potential~\cite{EMN} and zero for $c_D$ and $c_E$. The widths of Gaussian $\{\sigma_i\}$ are 1.0 except for $\tilde{c}_1$. From analyses on $c_1$ at NN-sector~\cite{EMN,EKMN}, we anticipated the $c_1$ having a smaller uncertainty, so the width is set $\sigma=0.5$.
We use g.s. energies of $^{40,48,52,54}$Ca and $^{56}$Ni as data to constrain the LECs, and the log-likelihood $L(\vec{c})$ is given with g.s. energies per nucleon
\begin{align}
L(\vec{c}) = -\frac{1}{2} \sum^{5}_{n=1}\left( \frac{E^\mathrm{exp.}_{n} - E^\mathrm{th.}_{n}(\vec{c})}{A_n}\right)^2\label{eq:llh},
\end{align}
where $E^\mathrm{exp.}$ is the experimental data, and $ E^\mathrm{th.}$ is the calculated result with given LECs, $\vec{c}$.

In step (ii),~we adopted the so-called affine invariant MCMC method~\cite{GW10} as the MCMC sampler and HF-MBPT as the many-body method to calculate g.s.~energies. The affine invariant MCMC method has several advantages: samples tend to show much shorter autocorrelation time compared to ordinary methods and one essentially does not need fine-tuning of the parameters in the sampler such as proposal distribution for random-walk Metropolis-Hastings, leap-flog parameters for Hamiltonian Monte Carlo, temperatures and their intervals for replica exchange methods, etc.
Since the sampler is an ensemble method, one can easily benefit from multi-node parallelizations.
We set the number of walkers, i.e. the number of nodes to be used, as 64 and generated $\geq 20,000$ samples after certain burn-in steps to confirm convergence and reproducibility by different runs.

\begin{figure}
\centering{
\includegraphics[width=\columnwidth]{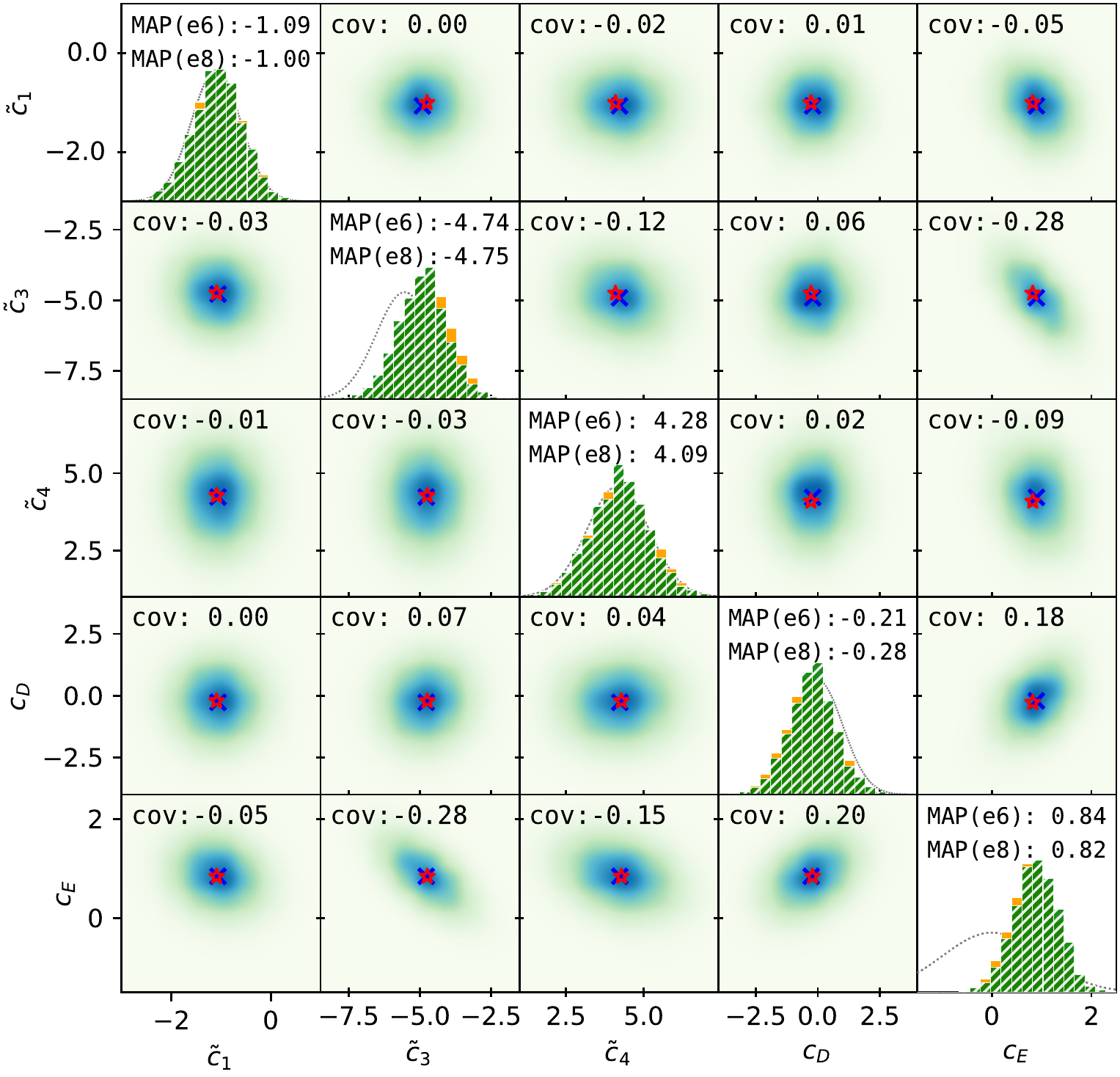}
\caption{MCMC samples over five LECs ($\tilde{c}_1, \tilde{c}_3, \tilde{c}_4, c_D$, $c_E$).
The lower (upper) non-diagonal panels show correlation plots of samples drawn from posteriors with $e_\mathrm{max}=6 (8)$ truncation, and the diagonal panels show histograms, where orange (hatched green) ones correspond to $e_\mathrm{max}=6 (8)$ results. The dotted lines show the prior distribution.
The cross (blue) and star (red) symbols are mean and MAP, respectively.\label{fig:LECs}}
}
\end{figure}

Regarding (iii), we plot the MCMC samples over five-dimensional LECs space in FIG.~\ref{fig:LECs}.
The panels in the lower and upper triangle, respectively, show density plots of the MCMC samples for $e_\mathrm{max}=6$ and $8$ and covariance between the two LECs.
The cross and star symbols show the mean values and the {\it maximum a posteriori} (MAP) values, respectively.
The MAP values are now evaluated by the parameters having the largest posterior ($\propto $ prior $\times$ likelihood) among the MCMC samples.
The diagonal panels show the histograms of MCMC samples for $e_\mathrm{max}=6(8)$ results and the MAP values.
Under our settings (2n3n, prior, data, etc.), $\tilde{c}_1$ shows almost no correlations with the others, whereas $c_E$ show rather modest correlations with $\tilde{c}_3,\tilde{c}_4,c_D$.
Note that, from our analysis, $\tilde{c}_1=-1.0, \tilde{c}_3 = -0.53, \tilde{c}_4 =-1.45, c_D = 1.0, c_E=-0.25$ gives a similar size of repulsion in terms of the g.s. energies of the target nuclei.
To take account of model space convergence, which is typically seen at $e_\mathrm{max}\sim 12$-$14$ for ${}^{40}$Ca with $\lambda_\mathrm{SRG} \leq 2.0 $ fm$^{-1}$, optimal values for the lower $pf$-shell region at $e_\mathrm{max}=8$ need to be modified so as to give a bit more repulsive results to the experimental values. 
Indeed, the MAP values at $e_\mathrm{max}=8$ give more repulsive results than those by the MAP at $e_\mathrm{max}=6$.
Since the $\tilde{c}_1$ is now uncorrelated to the others, we impose a role to gain such repulsion on $\tilde{c}_1$. It should be noted that this procedure can be valid only when adopting the current approximation and g.s. energies of $^{40,48,52,54}$Ca and $^{56}$Ni.

\begin{table}
 \caption{The parameter sets to be used in this study. The MAP* is constructed from maximum a posteriori (MAP) at $e_\mathrm{max}=8$ of the MCMC samples, and $\tilde{c}_1$ is slightly shifted to the larger value to gain a repulsion. The other ones will be used in IMSRG/VS-IMSRG calculation to see some possible difference in the nuclear structure.\label{tab:LECs}}
 \centering
 \begin{ruledtabular}
 \begin{tabular}{lcccccc}  
LECs & & $\tilde{c}_1$ & $\tilde{c}_3$ & $\tilde{c}_4$ & $c_D$  & $c_E$\\
     \hline
MAP* & & -1.100 & -4.753 & 4.089 & -0.277 & 0.820\\
 A & & -1.100 & -3.736& 4.514& -0.122& 0.344\\
 B & &  -1.600& -6.205& 3.326& -0.901& 1.520 \\
 C & & -0.950 & -3.415 & 4.594 & -1.400 & -0.182
\end{tabular}
\end{ruledtabular}
\end{table}

\begin{figure*}
\centering{
\includegraphics[width=18cm]{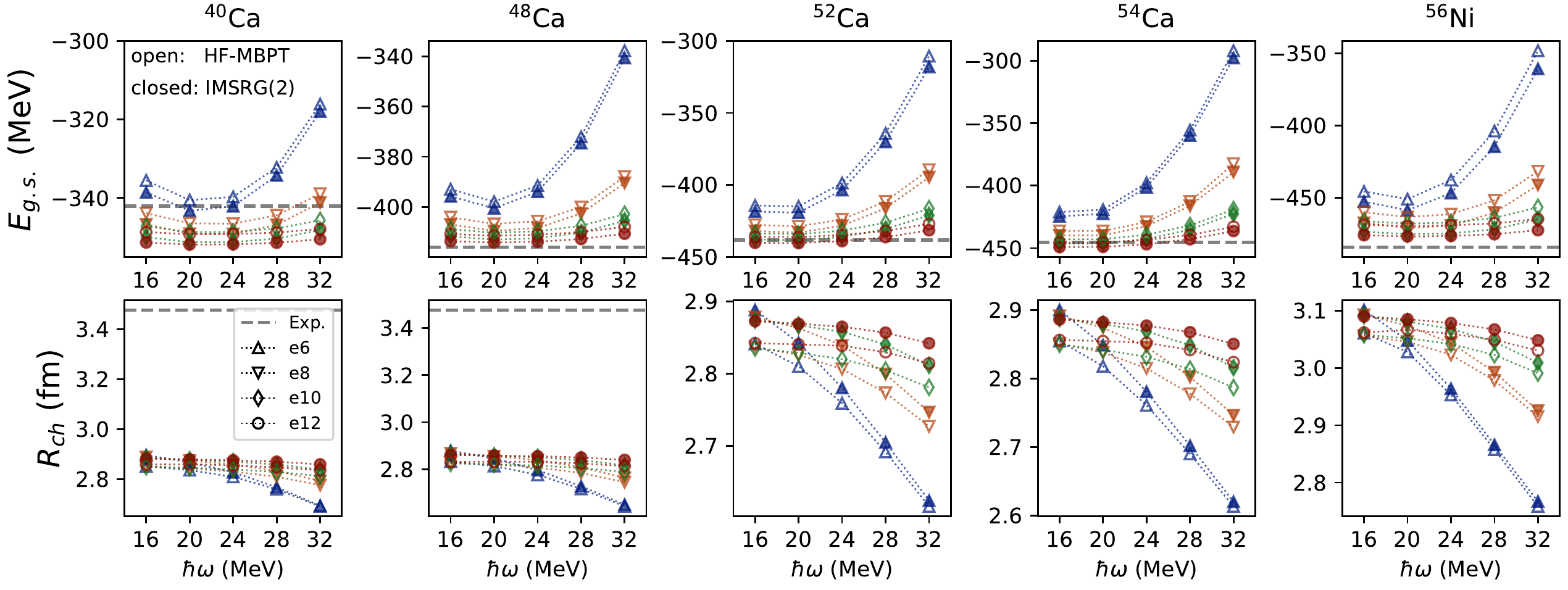}
\caption{Results with MAP* and convergence pattern of g.s. energies and charge radii with respect to model space size and oscillator parameter.
The open and filled symbols correspond to HF-MBPT and IMSRG(2) results, respectively.
The model space sizes, $e_\mathrm{max}=6,8,10,12$, are shown by different symbols.
\label{fig:conv}}
}
\end{figure*}

In this way, we constructed a tentative MAP value for larger $e_\mathrm{max}$, which will be called MAP* in the followings. In addition to the MAP*, we will also consider three-different parameters generated from the posterior distributions whose $\tilde{c}_1$ value is slightly changed for the reason above.
The aim to consider these parameters is to see the effect on the nuclear structure by some degrees of freedom, which may not be well constrained by only the g.s. energies. We list them in Table~\ref{tab:LECs}.



\paragraph*{Ground state properties with HF-MBPT/IMSRG(2).}

We discuss ground state properties, energy and charge radius, of sub-shell closed nuclei, ${}^{40,48,52,54}$Ca and ${}^{56}$Ni.
In FIG.~\ref{fig:conv}, the convergence pattern of g.s. energies and charge radii with respect to the $e_\mathrm{max}$ truncation and the oscillator parameter $\hbar\omega$ are shown.
The open and filled symbols correspond to HF-MBPT and IMSRG(2) results, respectively.
Note that the HF-MBPT is the 3rd (2nd) order for g.s. energies (charge radii), and more concrete expressions for expectation values for scaler operators can be found in, e.g.,~\cite{Miyagi22Sn}.
As a whole, the results are nearly converged at $e_\mathrm{max}=12$ and show rather flat behavior around $\hbar\omega = 20$ MeV, so we will use interactions with $e_\mathrm{max}=12, \hbar\omega=20$ MeV in the followings.

\begin{table}
 \caption{The IMSRG(2) results of g.s. energies using the four parameters with $\hbar\omega=20$ MeV, $e_\mathrm{max}=12$. \label{tab:gsEIMSRG}}
\centering
\begin{ruledtabular}
\begin{tabular}{lcccccc}
 $E_\mathrm{g.s.}$ (MeV)& & ${}^{40}$Ca & ${}^{48}$Ca & ${}^{52}$Ca & ${}^{54}$Ca & ${}^{56}$Ni \\
 \hline
MAP* & & -351.855 & -414.203 & -440.373 & -449.032 & -476.644 \\  
A & & -349.161 &  -414.154 & -443.224 & -453.427 & -476.886 \\
B & & -351.703 & -409.548 & -430.634 & -436.610 & -471.810 \\ 
C & & -344.348& -409.784  & -439.942 & -450.746 & -472.148 \\
Exp. & & -342.052 & -416.001 & -438.328 & -445.365 & -483.996   
\end{tabular}
\end{ruledtabular}
 \caption{The IMSRG(2) results of charge radii using the four parameters with $\hbar\omega=20$ MeV, $e_\mathrm{max}=12$. \label{tab:gsRIMSRG}}
\centering
\begin{ruledtabular}
\begin{tabular}{lcccccc}
$R_\mathrm{ch}$ (fm)& & ${}^{40}$Ca & ${}^{48}$Ca & ${}^{52}$Ca & ${}^{54}$Ca & ${}^{56}$Ni \\
 \hline
MAP* & & 2.8796 & 2.8582 & 2.8689 & 2.8823 & 3.0851\\
A &  & 2.8631 &  2.8350 & 2.8421 &  2.8539 & 3.0587\\
B & & 2.9068 & 2.8934 & 2.9100 & 2.9263 & 3.1209\\
C & & 2.8608 & 2.8300 & 2.8353 &  2.8464 & 3.0516 \\
Exp. & & 3.4776 & 3.4771 & -  & - & -
\end{tabular}
\end{ruledtabular}
\end{table}

In Tables~\ref{tab:gsEIMSRG} and \ref{tab:gsRIMSRG}, we summarized IMSRG(2) results for ground state properties of the target nuclei.
The g.s. energies by our interactions are well, but systematic underestimation is seen in charge radii.
This is consistent with the trade-off between binding energies and charge radii seen in previous studies, e.g.,~\cite{HoppeIMSRG}, but underestimation by our 2n3n interactions are more severe and almost irrelevant to the choice of LECs.

\begin{figure}
\centering{
\includegraphics[width=\columnwidth]{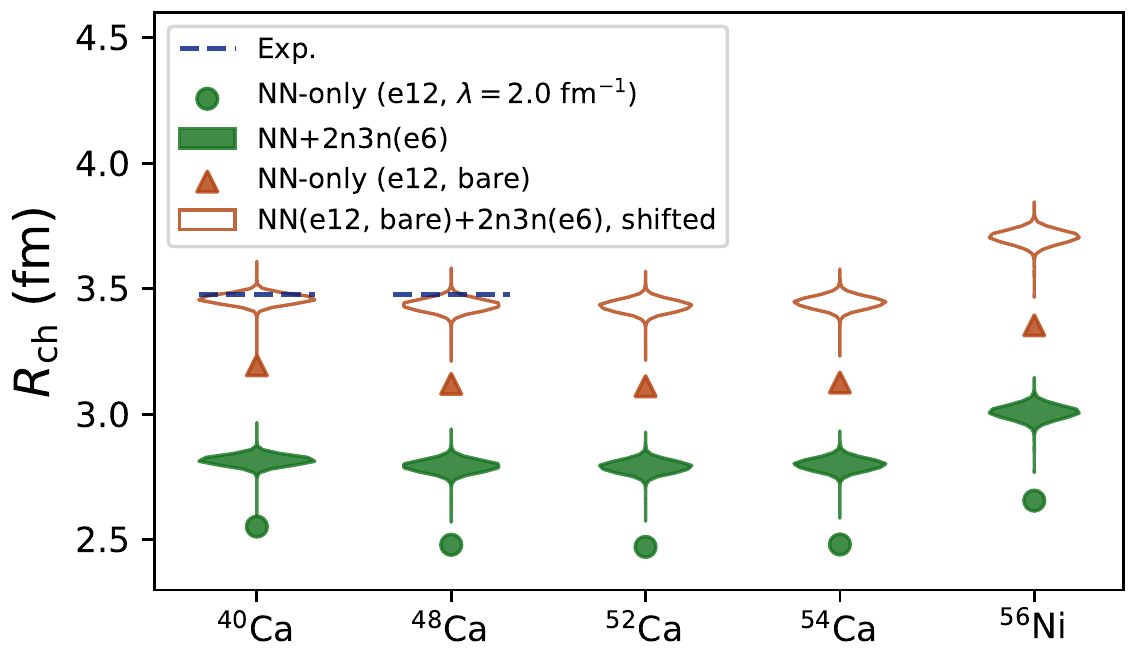}
\caption{Charge radii of the target nuclei. The NN-only results with $e_\mathrm{max}$ and different SRG resolution scales, $\lambda = 2.0 $ $\mathrm{fm}^{-1}$ and $\lambda=\infty$ (unevolved) are show by circles and triangles, respectively.
The closed violin plots show uncertainties of 2n3n contributions, which are evaluated by the MCMC samples ($e_\mathrm{max}=6$) shown FIG.~\ref{fig:LECs}, and the open violin plots show a naive estimates on charge radii by NN (bare) + 2n3n. See the main text for more details.\label{fig:uncertaintyRch}}
}
\end{figure}

Now let us look at possible origins of the underestimation from an uncertainty analysis.
We perform HF-MBPT(2) evaluation of charge radii for 10,000 MCMC samples with $e_\mathrm{max}=6$ and $\hbar\omega=20$ MeV, and the results are drawn as the violin plot (filled one) in FIG.~\ref{fig:uncertaintyRch}.
As shown in FIG.~\ref{fig:conv}, charge radii for ${}^{40,48}$Ca are almost converged at $e_\mathrm{max}=6$ and close to IMSRG(2) results, so only the HF-MBPT results with a smaller model space is enough to estimate the possible 2n3n contributions on charge radii.
In addition to this, we show the NN-only results with
SRG evolved interaction ($\lambda_\mathrm{SRG}=2.0$ fm${}^{-1}$,$e_\mathrm{max}=12$, circle symbols) and unevolved one ($e_\mathrm{max}=12$, triangle symbols).
From these, we found that there is no room for the 2n3n contribution, giving decent binding energies,\ to explain the observed charge radii.

The open violin plots show naive estimates of results under the bare NN plus 2n3n;
the open violins are drawn merely by shifting the closed violins by offsets,
$
R_\mathrm{ch}(\mathrm{e12}, \mathrm{bare})-
R_\mathrm{ch}(\mathrm{e12}, \lambda=2.0\ \mathrm{fm}^{-1} )
$.
This offset roughly shows the size of induced many-body forces, which are now ignored for brevity.
If one employs the harder NN interaction, the size of 2n3n needed to explain binding energies must be smaller, and thereby the 2n3n contributions on charge radii are to be smaller too. In this sense, the shifted values (open violin plots) in FIG.~\ref{fig:conv} may give us upper values by effective NN potentials.

We also note that, in the current approximation, all the contributions are truncated up to the two-body level by definition and there is no explicit normal ordered one-body (NO1B) contribution from both induced and genuine 3NF. This can deteriorate the consistent description of binding energies and charge radii.


Now, we explore the extensibility of the 2n3n approximation over the nuclear chart.
As already mentioned, density-dependent interactions have a dependence on Fermi momentum $k_F$ and we fixed it as the momentum from an empirical normal nuclear density.
Performing HF-MBPT calculation for ${}^{4}$He, ${}^{16}$O, ${}^{90}$Zr, and ${}^{132}$Sn, we naively estimate the validity of 2n3n interaction with a common $k_F$ in different mass regions.
For lighter nuclei, ${}^{4}$He and ${}^{16}$O, the binding energies are underestimated by about $10\%$, whereas the binding energies are overestimated for heavier ones, ${}^{90}$Zr and ${}^{132}$Sn, by about $10\%$. Regarding the slower convergence for heavier nuclei, fully converged results with respect to the model space size will be worse for ${}^{90}$Zr and ${}^{132}$Sn.
We also note that the charge radii are systematically underestimated as in the Ca and Ni regions, probably due to the same reasons as above.
From this observation, it is not straightforward to use the 2n3n approximation with a common $k_F$ in different mass ranges. One possible improvement is to introduce a phenomenological parameter to scale the 2n3n effect as a function of the mass number and to introduce a correction of the isospin asymmetry~\cite{HoltKaiserWeisse}, but this is left as a future study in a different context.


\paragraph*{Valence space results of lower $pf$-shell nuclei.}

\begin{figure}
\centering{
\includegraphics[width=\columnwidth]{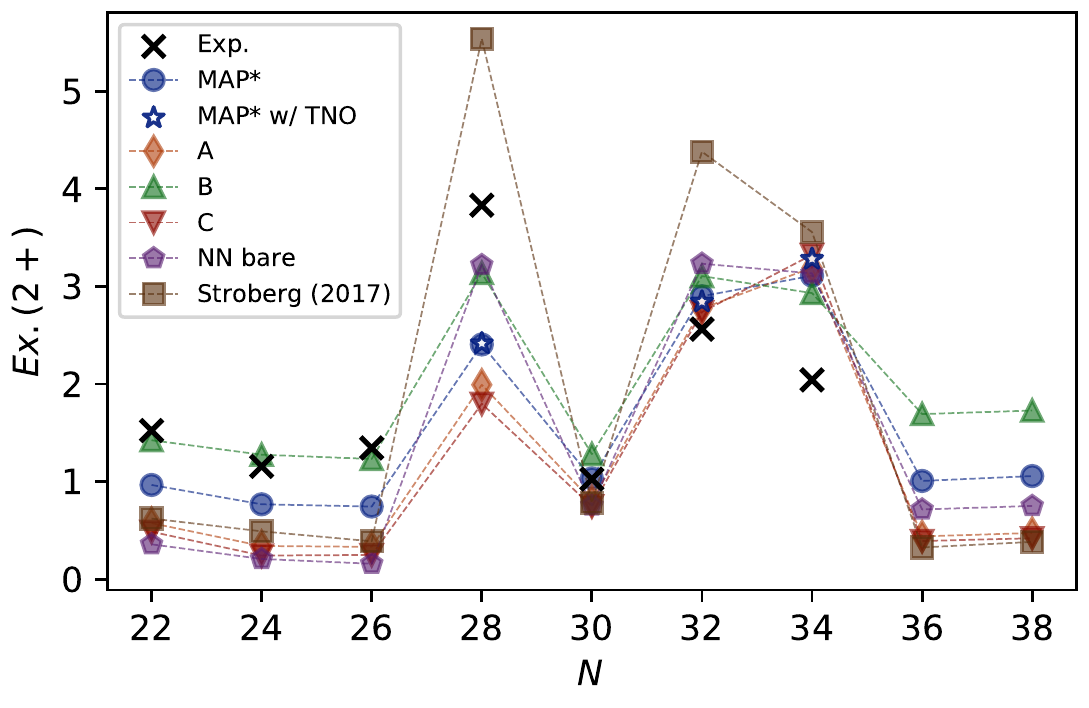}
\caption{
Systematics of first 2+ excitation energies of even calcium isotopes calculated by the VS-ISMRG.
The star and square symbols show the results with TNO or ENO, and the others are using effective interactions derived with the ${}^{40}$Ca reference. 
\label{fig:CaEx2}}
}
\end{figure}

Next, we discuss the shell-model results for some lower $pf$-shell nuclei using the effective interactions derived by the VS-IMSRG from our effective NN potentials.
We focus on the first 2+ energies for Ca isotopes, and low-lying spectra for ${}^{56}$Ni.

In FIG.~\ref{fig:CaEx2}, systematics of the first 2+ excited states of even calcium isotopes are shown for various different effective interactions.
The model space for the effective interactions is the $pf$ shell on top of the ${}^{40}$Ca core, and the reference state is ${}^{40}$Ca for our interactions, MAP*, A, B, and C.
As partly shown by open star symbols, the dependence on the reference state, which is examined using target normal ordering (TNO)/ensemble normal ordering (ENO)~\cite{StrobergRev19}, is minor for calcium isotopes.
One can see that most of the interactions give too low excitation energies especially for non-magic Ca isotopes.
The $N=28$ ($N=34$) gap is underestimated (overestimated) as a whole; only the exception is the "Stroberg2017" result at $N=28$. The interaction "Stroberg2017" is derived by the VS-IMSRG approach with consistently SRG evolved EM NN interaction at N3LO plus 3NF at NNLO, and the interaction files are available on the author's GitHub repository~\cite{Stroberg2017}.

From the trend seen in FIG.~\ref{fig:CaEx2}, the systematic underestimation of the first 2+ energies for non-magic Ca isotopes indicates that pairing components in $0f7/2$ and $0f5/2$ in the derived interactions are weak in general. Let us look at the results with an unevolved N4LO NN interaction, the pentagon symbols in the figure. While  the differences in the 3NF part are reflected as differences in 2+ energies, the results are essentially the same as those by the unevolved NN potential.
In this sense, one of the origins for weak pairing components is the NN interaction itself, although it is nontrivial to identify the origins of such trends and results are consequences of non-perturbative IMSRG/VS-IMSRG flow under given potentials. 

\begin{figure}
\centering{
\includegraphics[width=\columnwidth]{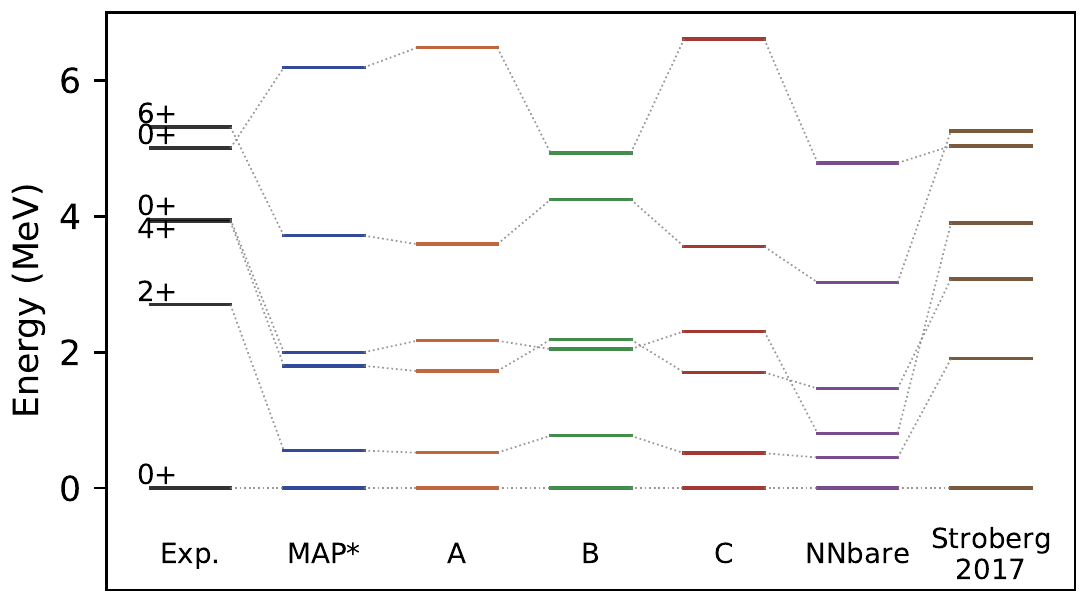}
\caption{Low-lying states of ${}^{56}$Ni by the VS-IMSRG approach.
The effective interactions are the same as in FIG.~\ref{fig:CaEx2}.
\label{fig:Ni56}}
}
\end{figure}

Next, we show the shell-model results of ${}^{56}$Ni in FIG.~\ref{fig:Ni56}.
All the effective interactions are derived for the $pf$ shell on top of the ${}^{40}$Ca core using TNO and diagonalized within the full $pf$-shell space. 
As can be expected from the results of ${}^{48}$Ca, the $N(Z)=28$ gap is underestimated by our effective interactions, leading to severe underestimation of the first 2+ energy.
Indeed, all the VS-IMSRG results with our interaction giving Ex.$(2+)\sim 500$ keV show that the occupation numbers of proton/neutron $0f7/2$ orbit are about $4$-$5$ for the $0^+_1$ and $2^+_1$ states.
The VS-IMSRG result by Ref.~\cite{Stroberg2017} gives a better description of the spectrum.
The effective interaction gives a rather prominent effective $N(Z)=28$ gap as partly shown ${}^{48}$Ca results above. This is partly due to the large gap between $0f7/2$ and $0p3/2$, $\sim 5-6$ MeV, in the single particle energies.

It should be noted that one can find the VS-IMSRG results with the so-called "EM 1.8/2.0" interaction~\cite{EM1820}, giving the 2+ systematics of Ca and Ni isotopes well described~\cite{SimonisIMSRG} and slightly overestimated $N(Z)=28$ gap.
The possible deficiency of pairing components and $N(Z)=28$ gap in NN potential discussed above would be compensated by the genuine 3NF,
which may lead to various successful descriptions by the EM1.8/2.0 interaction.
The systematic overestimation of 2+ energies of Ni isotopes by the VS-IMSRG with the EM1.8/2.0 is attributed to higher order contributions of IMSRG, and then the IMSRG(3) truncation has begun to be considered with lighter nuclei~\cite{Heinz2021}.

The $N(Z)=28$ gap in the VS-IMSRG approach is certainly related to the many-body machinery,
but it is also likely to be related to the absence of an explicit NO1B term from 3NF in effective NN force.
It is desirable to consider this point from more diverse perspectives.

\paragraph*{Summary and outlook.}

In this work, we explored the capability of chiral effective nucleon-nucleon (NN) potentials without introducing explicit three-nucleon force (3NF).
We employed a modern chiral NN interaction up to N4LO~\cite{EKMN,EMN} and a density-dependent interaction~\cite{HoltKaiserWeisse,Kohno,*KohnoErratum}, and calibrated the low-energy constants (LECs) for the density-dependent NN interaction by MCMC samplings.
We found the binding energies of some lower $pf$-shell nuclei are well explained by our effective NN potentials, whereas systematic underestimates were seen in charge radii and that is almost irrelevant to the choice of the LECs.
From an uncertainty analysis on the LECs and the results with a bare NN interaction, we discussed some possible origins of such underestimations.
From the VS-IMSRG results, significant underestimation is seen in the pairing of the $0f$ orbits and the $N(Z)=28$ gap.
The deficiency in the current effective interactions may be caused by a combination of multiple factors, including the NN potential itself, density-dependent approximation of 3NF, free-space SRG and the treatment of induced many-body forces, many-body methods, etc. 
For example, if we extend the model space to $pf$-$sdg$ shell in the VS-IMSRG decoupling phase, the 2+ energies of Ca isotopes and ${}^{56}$Ni improve by a few hundred keV.
The same trend can be seen in previous work using MBPT~\cite{HoltCa} and when one employs the extended Kuo-Krenciglowa method~\cite{NTsunoda2017} to derive interactions in the $pf$-$sdg$ space and diagonalizes it within the $pf$ shell.
This indicates that the inclusion of the $0g9/2$ orbit or others in the valence space can be important to derive effective interactions in this region.
The $pf$-shell (and beyond) nuclei would give us good testing grounds for these compound problems.

It should be also noted that sub-leading density-dependent NN forces in higher order have been already derived in Refs.~\cite{Kaiser2018,Kaiser2019,Kaiser2020}, and applications of these are left as future works along this study.

It is expected that this work and the package, NuclearToolkit.jl, will enable more detailed and multi-angle analyses on chiral potentials, many-body methods, and structure of medium-mass nuclei.

\paragraph*{Acknowledgments.}

The author thanks Noritaka Shimizu, Takayuki Miyagi, and Tokuro Fukui 
for discussions at the early stage of the development of NuclearToolkit.jl,
and Michio Kohno for discussions on the density-dependent NN force.
This work was supported by JSPS KAKENHI (Grants No. 22K14030) and partially by the computational resource of Fujitsu PRIMERGY CX400M1/CX2550M5 (Oakbridge-CX) at the Information Technology Center, The University of Tokyo.



\bibliography{p2n3n}

\end{document}